\documentclass[12pt]{article}

\newcommand{\beq}[1]{\begin{equation}\label{#1}}
\newcommand{\eeq}{\end{equation}}
\newcommand{\bear}[1]{\begin{eqnarray}\label{#1}}
\newcommand{\ear}{\end{eqnarray}}
\newcommand{\nn}{\nonumber}

\renewcommand{\theequation}{\arabic{section}.\arabic{equation}}
\catcode`\@=11 \@addtoreset{equation}{section}\catcode`\@=12

\newcommand{\np}{ {\newpage } }
\newcommand{\R}{\mbox{\bf R}}
\newcommand{\N}{ \mbox{\rm I$\!$N} }

\newcommand{\eps}{ \varepsilon }

\newcommand{\p}{\partial}
\newcommand{\btd}{\bigtriangledown}
\newcommand{\btu}{\bigtriangleup}
\newcommand{\tri}{\Delta}

\newcommand{\fnm}{\footnotemark}
\newcommand{\fnt}{\footnotetext}

\begin{document}

\begin{center}

\large \bf  Generalized
$pp$-wave solutions \\ on product of Ricci-flat
spaces

\end{center}

\vspace{1.03truecm}

\bigskip

\begin{center}

\normalsize\bf
V. D. Ivashchuk\fnm[1]\fnt[1]{e-mail: ivashchuk@mail.ru},

\bigskip

\it
 Center for Gravitation and Fundamental Metrology,
 SRCAM Rostest, 46 Ozyornaya ul., Moscow 119361, Russia

\it Institute of Gravitation and Cosmology,
Peoples' Friendship University of Russia,
6 Miklukho-Maklaya St., Moscow 117198, Russia

\end{center}

\begin{abstract}

A multidimensional gravitational model with
several scalar fields and form fields  is considered.
A wide class of generalized  $pp$-wave solutions
defined on a product of $n+1$ Ricci-flat spaces is obtained.
Certain examples of solutions (e.g. in supergravitational
theories) are singled out. For special cone-type internal factor spaces
the solutions are written in Brinkmann form.
An example of $pp$-wave solution is obtained using  Penrose limit
of a solution defined on a  product of two Einstein spaces.

\end{abstract}

\np

\section{\bf Introduction}
\setcounter{equation}{0}

Plane-fronted, parallel gravitational waves ($pp$-waves)
described by  the  metrics
 \beq{I.1}
 g = dx^{+} \otimes dx^{-}  +  dx^{-} \otimes dx^{+}
 + H(x^{-}, x) dx^{-} \otimes dx^{-} +
 \sum_{i = 1}^{m}   dx^{i} \otimes dx^{i},
 \eeq
with certain smooth functions $H(x^{-}, x)$
become rather popular objects of investigations
(see \cite{AmKl}-\cite{Ahm} and refs. therein).
The metric (\ref{I.1}) has a covariantly constant
null Killing vector $\p/\p x^{+}$ and obey the
vacuum Einstein equation when
 $\sum_{i = 1}^{m} \frac{\p^2}{\p x^i \p x^i} H = 0$.
In four dimensions it  is the most general solution to Einstein
equations  with covariantly constant null vector \cite{KSMH}.

The metrics (\ref{I.1}) also appeared
as exact solutions of string theory \cite{AmKl,HorSt1,HorSt2}
(e.g. with  non-trivial dilaton and null
3-form backgrounds)
due to vanishing of all higher order terms
in string   ($\beta$--function) equations of motion.
Despite of the triviality
of all scalar invariants constructed on powers
of the Riemann tensor and its derivatives
the metric (\ref{I.1}) is in general singular \cite{BGS,MR}.
For  $H(x^{-}, x) = A_{ij}(x^{-}) x^i x^j$ the metric is regular.
In the case of constant  $A_{ij}$ we are
led to so called Cahen-Wallach (CW) spaces \cite{CW}.
These spaces contain
maximally supersymmetric plane waves in
eleven-dimensional  \cite{KowGl,FP}
and ten-dimensional type $IIB$ \cite{BFHP} supergravity (see also
\cite{FP1}).  It was observed in \cite{BFHP:0201,BFP} that these solutions
may be also obtained as Penrose limits \cite{Pen,Guv} of the $AdS_p \times
S^{D-p}$ type solutions.

It was shown in  \cite{Me,MT,RT} that superstring theory on $pp$-wave
background can be solved in the light-cone gauge.
In \cite{BMN}  a sector
of $N = 4$ super-Yang-Mills   dual to string theory on
certain $pp$-wave background was identified (see also \cite{Ahm}),
allowing for stringy tests of the $AdS/CFT$ correspondence.

For $H(x^{-}, x) = A_{ij}(x^{-}) x^i x^j$
the solution (\ref{I.1}) may be also rewritten
in Rosen coordinates \cite{Guv,BFP,BL}
 \beq{I.2}
 g = du \otimes dv  +  dv \otimes du
 +  \sum_{i = 1}^{m} C_{ij}(u)  dy^{i} \otimes dy^{j}.
 \eeq

For diagonal matrix $(C_{ij}(u))= (C_{i}(u) \delta_{ij})$
the metric has a natural generalization
to a chain of Ricci-flat spaces.
In this paper we obtain a rather general class of solutions
defined on product of $n +1$ Ricci-flat spaces  for the multidimensional
gravitational model with fields of forms and scalar fields.
These solutions follow just from the equations of motion.

The paper is organized as follows.
In Sect. 2 we outline the general approach with arbitrary forms
and dilaton fields on a product of $(n+1)$ manifolds.
In Sect. 3 we obtain $pp$-wave  solutions defined on
a product of $(n+1)$  Ricci-flat spaces.
Here the integrability problem is also discussed.
Several examples of solutions are presented in Section 4.
In Section 5 a (generalized) Brinkmann form of solution for special
(cone-type) Ricci-flat factor-spaces is suggested.
Here an example of $pp$-wave solution is obtained using  Penrose limit
of a solution defined on product of two Einstein spaces \cite{I2}.

\section{\bf The model}
\setcounter{equation}{0}

We consider the model governed by the action
\bear{2.1}
S =&& \int_{M} d^{D}z \sqrt{|g|} \{ {R}[g] - 2 \Lambda - h_{\alpha\beta}
g^{MN} \partial_{M} \varphi^\alpha \partial_{N} \varphi^\beta
\\ \nn
&& - \sum_{a \in \Delta}
\frac{\theta_a}{n_a!} \exp[ 2 \lambda_{a} (\varphi) ] (F^a)^2 \},
\ear
where $g = g_{MN} dz^{M} \otimes dz^{N}$ is the metric,
$\varphi=(\varphi^\alpha) \in \R^l$
is a vector from dilatonic scalar fields,
$(h_{\alpha\beta})$ is a non-degenerate
symmetric $l\times l$ matrix ($l\in \N$), $\theta_a \neq 0$,
\beq{2.2}
F^a =  dA^a =\frac{1}{n_a!} F^a_{M_1
\ldots M_{n_a}} dz^{M_1} \wedge \ldots \wedge dz^{M_{n_a}}
\eeq
is a  $n_a$-form ($n_a \geq 2$) on a $D$-dimensional manifold $M$,
$\Lambda$ is
a cosmological constant and $\lambda_{a}$ is a $1$-form on $\R^l$ :
$\lambda_{a} (\varphi) =\lambda_{a \alpha}\varphi^\alpha$,
$a \in \Delta$; $\alpha=1,\ldots,l$.
In (\ref{2.1})
we denote $|g| = |\det (g_{MN})|$,
\beq{2.3}
(F^a)^2 =
F^a_{M_1 \ldots M_{n_a}} F^a_{N_1 \ldots N_{n_a}}
g^{M_1 N_1} \ldots g^{M_{n_a} N_{n_a}},
\eeq
$a \in \Delta$, where $\Delta$ is some finite set.
In the models with one time all $\theta_a =  1$
when the signature of the metric is $(-1,+1, \ldots, +1)$.

The equations of motion corresponding to  (\ref{2.1}) have the following
form
\bear{2.4}
R_{MN}  =   Z_{MN} + \frac{2\Lambda}{D - 2} g_{MN},
\\
\label{2.5}
{\btu}[g] \varphi^\alpha -
\sum_{a \in \Delta} \theta_a  \frac{\lambda^{\alpha}_a}{n_a!}
e^{2 \lambda_{a}(\varphi)} (F^a)^2 = 0,
\\
\label{2.6}
\nabla_{M_1}[g] (e^{2 \lambda_{a}(\varphi)}
F^{a, M_1 \ldots M_{n_a}})  =  0,
\ear
$a \in \Delta$; $\alpha=1,\ldots,l$.
In (\ref{2.5}) $\lambda^{\alpha}_{a} = h^{\alpha \beta}
\lambda_{\beta a}$, where $(h^{\alpha \beta})$
is a matrix inverse to $(h_{\alpha \beta})$.
In (\ref{2.4})
\beq{2.7a}
Z_{MN}= Z_{MN}[\varphi] +
\sum_{a \in \Delta} \theta_a e^{2 \lambda_{a}(\varphi)} Z_{MN}[F^a,g],
\eeq
where
\bear{2.7}
Z_{MN}[\varphi] =
h_{\alpha\beta} \p_{M} \varphi^{\alpha} \p_{N} \varphi^{\beta},
\\  \label{2.8}
Z_{MN}[F^a,g] = \frac{1}{n_{a}!}  \left[ \frac{n_a -1}{2 -D}
g_{MN} (F^{a})^{2}
 + n_{a}  F^{a}_{M M_2 \ldots M_{n_a}} F_{N}^{a, M_2 \ldots M_{n_a}}
 \right].
\ear

In (\ref{2.5}) and (\ref{2.6}) ${\btu}[g]$ and ${\btd}[g]$
are Laplace-Beltrami and covariant derivative operators respectively
corresponding to  $g$.

{\bf  Multi-index notations.} Let us consider the manifold
\beq{1.1}
M = M_{0}  \times M_{1} \times \ldots \times M_{n}.
\eeq
We denote $d_{i} = {\rm dim} M_{i} \geq 1$; $i = 0, \ldots, n$.
$D =  \sum_{i = 0}^{n} d_{i}$.
Let $g^i  = g^{i}_{m_{i} n_{i}}(y_i) dy_i^{m_{i}} \otimes dy_i^{n_{i}}$
be a metric on the manifold $M_{i}$, $i=1,\ldots,n$.

Here we use the notations of our previous papers
\cite{IMtop}.
Let any manifold $M_{i}$  be oriented and connected.
Then the volume $d_i$-form
\beq{2.A}
\tau_i  = {\rm dvol}(g^i) \equiv \sqrt{|g^i(y_i)|}
\ dy_i^{1} \wedge \ldots \wedge dy_i^{d_i},
\eeq
and the signature parameter
\beq{1.5}
\varepsilon_{i}  \equiv {\rm sign}( \det (g^i_{m_i n_i})) = \pm 1
\eeq
are correctly defined for all $i = 1,\ldots,n$.

Let $\Omega $  be a set of all non-empty
subsets of $\{ 1, \ldots,n \}$.
For any $I = \{ i_1, \ldots, i_k \} \in \Omega$, $i_1 < \ldots < i_k$,
we denote
\bear{1.6}
\tau(I) \equiv \hat{\tau}_{i_1}  \wedge \ldots \wedge \hat{\tau}_{i_k},
 \\ \label{1.6a}
\eps(I) \equiv \eps_{i_1}  \times  \ldots \times \eps_{i_k},
 \\ \label{1.7}
M_{I} \equiv M_{i_1}  \times  \ldots \times M_{i_k},
\\ \label{l.8}
d(I) \equiv  \sum_{i \in I} d_i,
\ear
where $d_i$ is both, the dimension of the oriented manifold $M_i$
and the rank of the volume form $\tau_i$ and $\hat{\tau}_i$
is the pullback of $\tau_i$ to the manifold $M$:
$\hat{\tau}_i = p_i^{*} \tau_i$, where
$p_{i} : M \rightarrow  M_{i}$, is the canonical projection,
$i = 1, \ldots, n$.

We also denote by
\beq{1.9}
\delta_{I}^i= \sum_{j\in I} \delta_{j}^i
\eeq
the indicator of $i$ belonging
to $I$: $\delta_{I}^i =1$ for $i\in I$ and $\delta^{i}_{I}=0$ otherwise.

\section{General solutions}

\subsection{Solutions governed by one equation}

Let $M_0$ be an open domain in $\R^2$ equipped with a flat metric
$g^0 = du \otimes dv + dv \otimes du$
where $u, v$ are coordinates.

Let us consider a plane wave metric
on the manifold (\ref{1.1}) of  the following form
\beq{2.2a}
g= \hat{g}^0  + \sum_{i=1}^{n} \exp(2 \phi^i(u)) \hat{g}^i,
\eeq
where  $g^i$  is a metric on $M_{i}$,  $i=1,\ldots,n$.

Here and in what follows $\hat{g}^{i} = p_{i}^{*} g^{i}$,
is the pullback of the metric $g^{i}$  to the manifold  $M$ by the
canonical projection: $p_{i} : M \rightarrow  M_{i}$, $i = 0,
\ldots, n$.

The fields of forms and  scalar fields
are also chosen in the $u$-dependent form
\bear{2.9}
F^a = \sum_{I \in \Omega_{a}} d \Phi^{(a,I)}(u) \wedge  \tau(I),
\\ \label{2.10}
\varphi^{\alpha} = \varphi^{\alpha}(u) .
\ear
where
$\Omega_{a} \subset \Omega$ are  non-empty subsets,
satisfying the relations $d(I) = n_a -1$
for all $I \in \Omega_a$, $a \in \Delta$.

The substitution of fields from
(\ref{2.2a}), (\ref{2.9}) and
(\ref{2.10}) into equations of motion
(\ref{2.4})-(\ref{2.6}) lead to the following
relations
\bear{2.11}
 {\rm Ric}[g^i] = 0,
 \\
 \label{2.12}
 \Lambda = 0, \\
 \label{2.13}
 - \sum_{i=1}^{n} d_i [ \ddot \phi^i +  (\dot\phi^i)^2] =
 h_{\alpha\beta}\dot\varphi^{\alpha}\dot\varphi^{\beta}
 + \sum_{s \in S} \eps_s \exp[ - 2 U^s(\phi,\varphi)] (\dot\Phi^s)^2
\ear
$i=1,\ldots,n$. In derivation of eqs.
(\ref{2.11})-(\ref{2.13}) the formulas for Ricci-tensor
and $Z$-tensor (\ref{2.7a}) from Appendix were used.

Here and in what follows,
${\rm Ric}[g^i]$  is the Ricci-tensor corresponding to $g^{i}$,
$i=1,\ldots,n$, and $\dot X \equiv dX/du$.
The (brane) index set  $S$ consists of elements
$s=(a_s,I_s)$, where $a_s \in \tri$ and
$I_s \in \Omega_{a_s}$ are ``color'' and
``brane'' indices, respectively.
The electric $U$-covectors and $\eps$-symbols
are defined as follows \cite{IMtop}
\bear{2.u}
  U^s = U^s(\phi,\varphi)=
  - \lambda_{a_s}(\varphi) + \sum_{i \in I_s} d_i \phi^i,
 \\ \label{2.e}
 \eps_s= \eps(I_s) \theta_{a_s}
\ear
for $s=(a_s,I_s) \in S$.

Thus, we get from equations of motion
that cosmological term is zero
and all spaces $(M_{i},g^i)$
($i=1,\ldots,n$) are Ricci-flat.
We also are led  to
the second order differential equation
(\ref{2.13}) on logarithms of scale-factors $\phi^i$,
scalar fields $\varphi^{\alpha}$ and
"brane" scalar fields $\Phi^s$.

The solution (\ref{2.11})-(\ref{2.13})
is valid without imposing of any  restrictions
(on brane configurations) or
intersection rules.
(Compare with the solutions from \cite{IMtop})
and references therein).

\subsubsection{ "Electro-magnetic" form of solution}

Now we show that more general composite
"electro-magnetic" ansatz
\beq{2.9a}
F^a = \sum_{I \in \Omega_{a,e}} d \Phi^{(a,e,I)}(u) \wedge  \tau(I)+
\sum_{J \in \Omega_{a,m}} e^{ - 2 \lambda_{a}(\varphi)}
* (d \Phi^{(a,m,J)}(u) \wedge  \tau(J)),
\eeq
(instead of composite electric one from (\ref{2.9}))
will not give new solutions. Here
$*=*[g]$ is the Hodge operator on $(M,g)$,
$\Omega_{a,e}, \Omega_{a,m} \subset \Omega$ are  non-empty subsets,
satisfying the relations:
$d(I) = n_a -1$ for all $I \in \Omega_{a,e}$ and
$d(J) = D- n_a -1$ for all $J \in \Omega_{a,m}$,
$a \in \Delta$.

Indeed,  due to relations   (5.26) from \cite{IM3}
we get
 \bear{2.24}
 * (d \Phi \wedge \tau(I)) =
                   P_I   (*_0 d \Phi) \wedge \tau(\bar{I})
                =  P_I   d \Phi \wedge \tau(\bar{I}),  \\
  \label{2.24a}
   P_{I} =  \eps(I) \mu(I) \exp[ \sum_{j = 1}^{n} d_j \phi^j
             - 2 \sum_{i \in I} d_i \phi^i ],
 \ear
 where  $\bar{I}  = \{1, \ldots,n \} \setminus I$
is "dual" set,  $\mu(I) = \pm 1$  is defined by
the formula
$\tau(\bar{I}) \wedge du \wedge \tau(I)=
\mu(I) du \wedge \tau(\{1, \ldots,n \})$ and
$*_0=*[g^0]$ is the Hodge operator on $(M_0,g^0)$,
obeying  $*_0 d \Phi = d \Phi$ for $\Phi = \Phi(u)$.
\fnm[2]\fnt[2]{Here we use the following definition
for Hodge dual:
$$(* \omega)_{M_1...M_{D-k}} = \frac{|g|^{1/2}}{k!}
\eps_{M_1...M_{D-k} N_{1}...N_{k}} \omega^{N_{1}...N_{k}},$$
where ${\rm rank} (\omega) = k$. We also put $\eps_{uv} =  1$.}

Using (\ref{2.24}) any "magnetic"
monom may be rewritten in the electric form
 \beq{2.25}
\exp( - 2 \lambda_{a}(\varphi) )
* (d \Phi^{(a,m,J)}) \wedge  \tau(J)) =
(d \bar{\Phi}^{(a,m,J)}) \wedge  \tau(\bar{J})),
\eeq
where
$d \bar{\Phi}^{(a,m,J)} = \exp( - 2 \lambda_{a}(\varphi(u)) )
P_J(u)  d \Phi^{(a,m,J)}$, $J \in \Omega_{a,m}$.

{\bf Remark.}
A rather simple way to verify that
fields of forms (\ref{2.9})
obey  the "Maxwell" equations
(\ref{2.6}) written in the form $d * (e^{2 \lambda_{a}(\varphi)}
F^{a})  =  0$ is to use the relation
(\ref{2.24}).

\subsection{The integrability problem}

Here  we consider the problem of integrability of
equation (\ref{2.13}) that may be rewritten in equivalent form
in terms of scale factors $f_i = \exp(\phi_i)$
as following
\bear{3.1}
- \sum_{i=1}^{n} d_i \frac{\ddot{f}_i}{f_i} =
h_{\alpha\beta}\dot\varphi^{\alpha}\dot\varphi^{\beta}
+ \sum_{s \in S} \eps_s \exp[ 2 \lambda_{a_s}(\varphi)] (\dot\Phi^s)^2
\prod_{i \in I_s} f_i^{-2d_i}.
\ear

Let us consider the following problem: for given
scale factors $f_2, \dots, f_n$, scalar fields $\varphi^\alpha$
and brane scalar fields $\Phi^s$ to find $f_1$
satisfying eq. ({\ref{3.1}}).
The solution of this problem is equivalent
to finding of all metrics (\ref{2.2}) obeying equations of motion
(\ref{2.4})-(\ref{2.6}) for given fields of forms (\ref{2.9}) and scalar
fields (\ref{2.10}).

{\bf Remark.} Here we do not  consider another
trivial task: for given
scale factors $f_1, \dots, f_n$ and scalar fields $\varphi^\alpha$
to find all brane scalar fields $\Phi^s$
satisfying eq. ({\ref{3.1}}).

Denoting $\Psi = f_1(u)$ we get the following
non-linear equation
\beq{3.2}
\ddot\Psi = A(u)\Psi + B(u) \Psi^{1 - 2 d_1},
\eeq
where function $A = A(u)$ and $B = B(u)$ are defined
as follows
\bear{3.3a}
A = - \frac{1}{d_1} \left[
\sum_{i=2}^{n} d_i \frac{\ddot{f}_i}{f_i} +
h_{\alpha\beta}\dot\varphi^{\alpha}\dot\varphi^{\beta} \right.
+ \\ \nn
\left.
\sum_{s \in S \setminus S_1} \eps_s \exp[ 2 \lambda_{a_s}(\varphi)]
(\dot\Phi^s)^2
\prod_{i \in I_s} f_i^{-2d_i} \right],
\\ \label{3.3b}
B = - \frac{1}{d_1}
\sum_{s \in S_1} \eps_s \exp[ 2 \lambda_{a_s}(\varphi)]
(\dot\Phi^s)^2
\prod_{i \in I_s \setminus \{1 \} } f_i^{-2d_i}.
\ear

Here we denote
\beq{3.4}
S_1 \equiv \{ s \in S: 1 \in I_s \}.
\eeq
The subset $S_1 \subset S$ describes all branes that "cover"
the space $M_1$.

A non-linear equation (\ref{3.2}) with $d_1 \neq 0,1$
was considered by Reid in \cite{Reid},
where  a special subclass of solutions
was obtained for certain  functions $B(u) = B(u,d_1)$.

\subsubsection{One factor-space is "free"}

Let us suppose that one of factor spaces, say
$M_1$, is not "occupied" by branes, i.e.
\beq{3.5}
1 \notin I_s
\eeq
for all $s \in S$.

In this case $B = 0$ and we get
a very familiar (from quantum mechanics) linear equation
\beq{3.2a}
\ddot\Psi = A(u) \Psi,
\eeq
where function $A = A(u)$ is  defined
is (\ref{3.3a}) with $S_1 = \emptyset$.
For certain "potentials" $A(u)$ one can
find explicit general solutions to
(Schroedinger-type) eq. (\ref{3.2a}).

\section{Special solutions}

\subsection{$Sin$-type solutions.}

Let us consider a special class of solutions
with $sin$-type dependent scale factors
\bear{4.1}
g= du \otimes dv + dv \otimes du +
\sum_{i=1}^{n} f_i^2 \hat{g}^i,
  \\ \label{4.2}
F^a = \sum_{s \in S}  \delta_a^{a_s} q_s
e^{ -\lambda_{a_s}(\varphi)}
(\prod_{i \in I_s} f_i^{d_i})
du \wedge  \tau(I_s),
\\ \label{4.3}
\varphi^{\alpha} = {\rm const} ,
\ear
where
\beq{4.1a}
f_i = c_i \sin( \omega_i u + \omega_i^0 ),
\eeq
$c_i > 0$, $\omega_i$, $\omega_i^0$ and $q_s$
are constants ($i=1,\ldots,n$, $s=(a_s,I_s) \in S$)
obeying  the following relation
\beq{4.4}
\sum_{i=1}^{n} d_i \omega_i^2 = \sum_{s \in S}  \eps_s q_s^2.
\eeq

These special solutions (with $\eps_s = +1$, $\theta_a =+1$)
have important applications  in supergravity and string theory.

In what follows we consider two examples of these
$sin$-type solutions.

\subsection{Solutions in $IIB$ supergravity.}

Let us consider a solution in $IIB$-supergravity
with two $4$-dimensional Ricci-flat spaces
$(M_i,g^i)$, $i = 1,2$, of Euclidean signature.

We consider a sector with 5-form and
put $\varphi = 0$.
We also put $I_1 = \{ 1 \}$, $I_2 = \{ 2 \}$;
$c_i =1$, $\omega_i = 1$, $\omega_i^0 = 0$
($i =1,2$) and $q_s = \pm 2$, $s \in \{ s_1, s_2 \}$.

In this case the solution (\ref{4.1}), (\ref{4.2}) reads
as follows
\bear{4.1c}
g= du \otimes dv + dv \otimes du +
\sum_{i=1}^{2}  \sin^2(u)  \hat{g}^i,
  \\ \label{4.2c}
F_{[5]} = \pm 2 (\sin^4(u)) [du \wedge \hat{\tau}_1 + du \wedge
\hat{\tau}_2],
\ear
with $F_{[5]}= * F_{[5]}$ (see  (\ref{2.24})).

For flat spaces $(M_i,g^i)$, $i = 1,2$, we get a
well-known  supersymmetric solution from \cite{BFHP},
written in the Rosen representation.

\subsection{Solutions in $D=11$ supergravity.}

Now we consider a solution in $D =11$-supergravity
\cite{CJS} with two Ricci-flat spaces
$(M_i,g^i)$, $i = 1,2$, of Euclidean signature
and dimensions $d_1 =3$ and $d_2 = 6$, respectively.
Here the first space $(M_1,g^1)$ is obviously flat,
since it is 3-dimensional  Ricci-flat space.

Let $I_1 = \{ 1 \}$
(i.e. one brane "living" on $M_1$ is considered);
$c_1 = 1/2$, $c_2 = 1$, $\omega_1 = 1$, $\omega_2 = 1/2$,
$\omega_i^0 = 0$ ($i =1,2$) and $q_s = \pm 3/\sqrt{2}$, $s = s_1$.

Then the solution (\ref{4.1}), (\ref{4.2}) reads
as follows
\bear{4.1b}
g= du \otimes dv + dv \otimes du +
\frac{1}{4}  \sin^2(u)  \hat{g}^1 +
\sin^2(\frac{1}{2} u)  \hat{g}^2,
  \\ \label{4.2b}
F_{[4]} = \pm \frac{3}{8\sqrt{2}} (\sin^3(u)) du \wedge \hat{\tau}_1,
\ear

For flat space $(M_2,g^2)$ we are led to
supersymmetric solution from \cite{FP},
written in the Rosen representation.

\subsection{Solutions with constant scale factors}

 A special class of solutions occurs when all
 scale factors are constant. In this case relation
 (\ref{2.13b}) implies

   \beq{2.13b}
 0 =
 h_{\alpha\beta}\dot\varphi^{\alpha}\dot\varphi^{\beta}
 + \sum_{s \in S} \eps_s \exp[ - 2 U^s(\phi,\varphi)] (\dot\Phi^s)^2.
   \eeq

 Usually, in all substantial (supergravitational) examples the matrix
 $(h_{\alpha\beta})$ is positive definite and  all $\eps_s$ are positive
 (for pseudo-Euclidean signature $(-,+, ..,+)$ of $g$). Hence, it follows
  from (\ref{2.13b}) that in this case
  all scalar fields are constant and all fields of
 forms are zero. But one may obtain non-trivial solutions when
 $(h_{\alpha\beta})$ is not positive definite one, or, when some of
 $\eps_s$ are negative. Such solutions occur in twelve dimensional model
 \cite{KKLP}, corresponding to $F$-theory and  in so-called $B_D$-models
 \cite{IMJ} in dimension $D \geq 12$.

\section{Solutions with special factor spaces}

\subsection{Brinkmann form of special solutions}

In this Section we show that
the solutions under consideration may be
written in generalized "Brinkmann form"
when  special "internal" factor spaces
are chosen.

Let $M_i = \R_{+} \times N_i$ and
 \beq{5.1}
     g^i = dr_i \otimes dr_i + r_{i}^2 \hat{h}_i,
 \eeq
where $(h_i, N_i)$ is Einstein space of dimension $d_i -1$
and ${\rm Ric}[h_i] = (d_i - 2) h_i$.
Clearly, all metrics (\ref{5.1}) are Ricci-flat.

For cone-type metrics (\ref{5.1})  the
relations (\ref{2.2a}) and (\ref{2.9})
may be rewritten as follows
\bear{5.2a}
g = du \otimes d \bar{v}  +  d \bar{v} \otimes d u
- \left(\sum_{i = 1}^{n} \lambda_i(u) r_i^2 \right) d u \otimes d u +
\sum_{i = 1}^{n}  \hat{g}^{i},
  \\ \label{5.2b}
F^a = \sum_{s \in S}  \delta_a^{a_s}
d \bar{\Phi}^s(u) \wedge  \tau(I_s),
\ear
and  eq. (\ref{2.13}) reads
\beq{5.4}
\sum_{i=1}^{n} d_i \lambda_i =
h_{\alpha\beta}\dot\varphi^{\alpha}\dot\varphi^{\beta}
+ \sum_{s \in S} \eps_s \exp[ 2 \lambda_{a_s}(\varphi)]
(\dot{\bar{\Phi}^s})^2.
\eeq

The metric (\ref{5.2a}) may be obtained
if one substitute into original metric  (\ref{2.2a})
the internal metric
$ g^i = dR_i \otimes dR_i + R_{i}^2 \hat{h}_i$, and
then make redefinition of coordinates
\bear{5.5}
\bar{v} = v + \frac12 \sum_{i = 1}^{n} h_i(u) R_i^2,
  \\ \label{5.6}
r_i = f_i(u)R_i, \quad h_i = - f_i \dot{f}_i,
\ear
$i = 1, \ldots, n$. In (\ref{5.4})
$\lambda_i = -  \ddot{f}_i/f_i$ and
  $\dot{\bar{\Phi}^s}  = \dot{\Phi}^s \prod_{i \in I_s } f_i^{-d_i}$,
  $i = 1, \ldots, n$; $s \in S$.

It should be noted that the class of supersymmetric
$pp$-wave solutions in ten dimensional $IIB$ supergravity
obtained by Maldacena and Maoz (see also \cite{NKim}) has a
non-empty intersection with the family of our solutions for
$D =10$.

\subsection{Penrose limit of a solution
on product of two Einstein spaces}

Some of $pp$-wave solutions may be obtained from
generalized Freund-Rubin-type solutions  \cite{I2} (defined on
product of Einstein spaces) using the Penrose limit. Here we consider an
example of such procedure.

Let us consider the special solution from  \cite{I2}
defined on product of two Einstein spaces
$(\bar{M}_i,\bar{g}_i)$,
\beq{A3.0}
{\rm Ric}(\bar{g}_i) = \xi_i \bar{g}_i,
\eeq
$i = 1,2$, with equal dimensions
$d_1 = d_2 = d + 2$.

The static Freund-Rubin-type solution
with one non-zero form reads
\bear{A3.1}
g = \hat{\bar{g}}_1 + \hat{\bar{g}}_2, \\
\label{A3.2}
F^a = Q_1 \hat{\tau}_1 + Q_2 \hat{\tau}_2, \\
\label{A3.3}
\varphi^{\alpha} = 0, \\
\label{A3.4}
\xi_1 = - Q^2, \qquad \xi_2 =  Q^2,
\ear
where $Q_1^2 = Q_2^2 =  Q^2$.

Let  $\bar{M}_i = T_i \times \R \times N_i$, where
$T_i \subset \R$ are intervals. We consider
Einstein metrics  on  $\bar{M}_1$ and $\bar{M}_2$,
respectively,
\bear{A3.5}
\bar{g}_1 = (d \rho \otimes d \rho - \cosh^2(\rho) dt \otimes dt
+  \sinh^2(\rho) \hat{h}_1) R^2,
\\ \label{A3.6}
\bar{g}_2 = (d \theta \otimes d \theta - \cos^2(\theta) d \psi \otimes
d \psi +  \sin^2(\theta) \hat{h}_2) R^2,
\ear
generated by  Einstein spaces $(h_i, N_i)$
(of dimension $d$) obeying ${\rm Ric}[h_i] = (d - 1) h_i$, $i = 1, 2$.
Here $R^2 = (d+2)/Q^2$.

Introducing new variables
\bear{A3.7}
 \bar{x}^{\mp} = (t \pm \psi)/2, \quad x^{-} = \bar{x}^{-}/\mu,
 \quad x^{+} = - \bar{x}^{+} 2 \mu R^2, \\
\label{A3.8}
 r_1 = \rho R, \qquad r_2 = \theta R,
\ear
where $\mu \neq 0$ and taking the Penrose limit $R \to + \infty$,
we get
\bear{A3.9}
g = d x^{+}  \otimes d x^{-}  +   d x^{-}  \otimes d x^{+}
- \mu^2 ( r_1^2 + r_2^2) d x^{-} \otimes d x^{-} +
 \hat{g}_1 + \hat{g}_2,
  \\ \label{A3.10}
F^a = \mu \sqrt{d+1} dx^{-} \wedge \left( \delta_1 \hat{\tau}_1
+ \delta_2 \hat{\tau}_2  \right),
\ear
where $\delta_i = \pm 1$  ($\delta_i = - {\rm sign} Q_i$),
$\tau_i = {\rm dvol}(g_i)$
and $ g_i = dr_i \otimes dr_i + r_{i}^2 \hat{h}_i$,
Ricci-flat metrics on $M_i = \R \times N_i$,  $i = 1,2$.

\section{Discussions}

In this paper we obtained  exact solutions
describing plane waves on
the product of $(n+1)$ Ricci-flat spaces
for the gravitational model with fields of forms and
(dilatonic) scalar fields. The solutions are
given by the relations (\ref{2.2a})-(\ref{2.13})
 and may be considered
as a composite  generalization of  well-known
$pp$-wave solutions in supergravitational theories.

The general solutions are defined up to solutions of
the second order differential equation  (\ref{2.13}).
When i) one of factor spaces is not "occupied" by branes then
the problem is reduced to Schroedinger-type eq. (\ref{3.2a}).
In the case ii) when all spaces are occupied by branes
we are led to non-linear equation (\ref{3.2}) (Reid equation).
An interesting question here is to find a possible
"chaotic"  behaviour among solutions of eq. (\ref{3.2}), as
it sometimes takes place in  "cosmology"  with
$p$-branes \cite{IMb,DHN}. In the first case i) the "chaotic"
behaviour seems to be absent for $pp$-wave solutions (see (\ref{3.2a}))
in agreement with Kasner-like (non-oscillating) behaviour
near the singularity for $S$-brane cosmology \cite{Is-b} (when one space
is "free").

Another interesting problem is related to classification of
(fractional) supersymmetric configurations among the solutions
under consideration. On this way the earlier results of the paper
\cite{Isusy}  may be used.

Here we considered as an example a solution from \cite{I2}
("Freund-Rubin" type solution) defined on product
of two Einstein spaces and showed
the appearance of a $pp$-wave solution in the Penrose limit.
An interesting problem
is to investigate all possible Penrose limits of the
composite p-branes solutions on product of Einstein spaces
 from \cite{I2} (we remind that certain solutions
 from \cite{I2} appear in the "near-horizon" limit
of solutions with harmonic functions and
Ricci-flat internal spaces \cite{IM3,IMtop}) .

 \begin{center}
  {\bf Acknowledgments}
 \end{center}

This work was supported in part
by the DFG grant  436 RUS 113/678/0-1(R),
by the Russian Ministry of
Science and Technology and  Russian Foundation for Basic Research,
grant 01-02-17312. The author thanks Prof. H. Nicolai for
hospitality during the  visit to Albert Einstein Institute (December 2002).
The author is also grateful to Prof. S. Theisen and Dr. N. Kim for useful
discussions.

\renewcommand{\theequation}{\Alph{subsection}.\arabic{equation}}
\renewcommand{\thesection}{}
\renewcommand{\thesubsection}{\Alph{subsection}}
\setcounter{section}{0}

\section{Appendix}

\subsection{Ricci and Riemann tensors}

The non-zero (identically) Ricci tensor components
for the metric (\ref{2.2a}) are the following
(see Appendix in \cite{IMtop})
\bear{A1.1}
R_{u u}[g]  = - \sum_{i=1}^{n} d_i [ \ddot \phi^i +  (\dot\phi^i)^2],
\\
\label{A1.2}
R_{m_{i} n_{i}}[g]  = {R_{m_{i} n_{i}}}[g^i],
     \ear
$i = 1, \dots, n$.

The scalar curvature for (\ref{2.2a}) reads
\bear{A1.3}
  R[g] =  \sum_{i =1}^{n} e^{-2 \phi^i} {R}[g^i].
\ear

The non-zero (identically) components of the Riemann tensor corresponding
to the metric  (\ref{2.2a})  have the following form
\bear{A1.5}
&&R_{u m_i u n_i }[g]    = - R_{m_i u u n_i }[g] =
- R_{u m_i n_i u }[g]    =  \nonumber \\
&& R_{m_i u n_i u }[g] =  - \exp(2 \phi^i) g^{i }_{ m_i n_i}
[ \ddot \phi^i +  (\dot\phi^i)^2],
\\     \label{A1.6}
&& R_{m_i n_i p_i q_i}[g]  = \exp(2 \phi^i)
 R_{m_i n_i p_i q_i }[g^i],
\ear
$i = 1, \dots, n$.

\subsection{Product of forms}

For two forms $F_1$ and $F_2$  of rank $r$ on $(M,g)$
($M$ is a manifold and
$g$ is a metric on it) we use notations
\bear{A2.1}
(F_1\cdot F_2)_{MN} \equiv
{(F_1)_{MM_2\dots M_r}(F_2)_N}^{M_2\cdots M_r},
\\ \label{A2.2}
F_1 F_2 \equiv (F_1\cdot F_2)_M^M  =
(F_1)_{M_1M_2\dots M_r}(F_2)^{M_1M_2\cdots M_r}.
\ear

It may be verified
(see also formulas in Appendix from
\cite{IMtop}) that
\beq{A2.14}
{\cal F}^{(a,I)} {\cal F}^{(a,J)}=0,
\eeq
$I,J \in \Omega_{a}$,
where
\beq{A2.14a}
{\cal F}^{(a,I)} = d \Phi^{(a,I)}(u) \wedge  \tau(I).
\eeq
Hence for  composite fields
\beq{A2.15}
F^{a}= \sum_{I \in \Omega_{a}} {\cal F}^{(a,I)},
\eeq
we get
\beq{A2.16}
(F^{a})^2= 0,
\eeq
$a \in \Delta$.

Let us consider the tensor
$F^{a}\cdot F^{a} = (F^{a}\cdot F^{a})_{MN} dz^M \otimes dz^N$
for composite $F^{a}$ from (\ref{A2.15})
One verify can that
\beq{A2.17}
({\cal F}^{(a,I)}\cdot{\cal F}^{(a,J)})_{MN} =0,
\eeq
for $I \neq J$, $I,J \in \Omega_{a}$, $a \in \Delta$.
This may done using the relations
 from Appendix of \cite{IMtop} for "non-dangerous"
intersection $d(I \cap J) \neq d(I) -1$ and
relation $(4.14)$ from \cite{IM3} for "dangerous"
intersection \cite{AR} $d(I \cap J) = d(I) -1$.

Now we put $I = J \in \Omega_{a}$.
The only non-zero (identically) components
of the tensor $({\cal F}^{(a,I)} \cdot {\cal F}^{(a,I)})$
are the following ones (see Appendix in \cite{IMtop})
\beq{A2.18}
 ({\cal F}^{(a,I)} \cdot {\cal F}^{(a,I)})_{uu} =
 (n_a -1)!
 \eps(I) \exp[ - 2 \sum_{i \in I} d_i \phi^i ]
 (\dot\Phi^{(a,I)})^2.
\eeq

It should be noted, that the key ingredient in verification
of formulas (\ref{A2.15})
and (\ref{A2.17}) (and some other
formulas from Section 2) is the following obvious relation
$g^{0, \mu \nu} \p_{\mu} \Phi \p_{\nu} \Phi^{'} = 0$
for 2-metric  $g^0 =  du \otimes dv + dv \otimes du$ and
functions $\Phi =  \Phi(u)$, $\Phi^{'} =  \Phi^{'}(u)$.

It follows from (\ref{A2.14}), (\ref{A2.17}) and (\ref{A2.18})
that the only non-zero (identically) components
of $Z$-tensor (\ref{2.7a}) read as follows
\beq{A2.7b}
 Z_{uu} =
 h_{\alpha\beta} \dot\varphi^{\alpha} \dot\varphi^{\beta}
 + \sum_{a \in \Delta} \theta_a
 \sum_{I \in \Omega_{a}} \eps(I)
 \exp[2 \lambda_{a}(\varphi) - 2 \sum_{i \in I} d_i \phi^i ]
 (\dot\Phi^{(a,I)})^2.
\eeq
This relation coincides with the right hand side of the
equation  (\ref{2.13}).


\small
\begin{thebibliography}{99}

\bibitem{AmKl}
D. Amati and C. Klimcik,
{\it Phys. Lett.} {\bf B 219}, 443 (1989).

\bibitem{HorSt1}
G.T. Horowitz and A.R. Steif,
{\it Phys. Rev. Lett.} {\bf 64},  260 (1990).

\bibitem{HorSt2}
G. T. Horowitz and A.R. Steif,
{\it Phys. Rev.} {\bf D 42}, 1950 (1990).

\bibitem{KowGl}
J. Kowalski-Glikman,
{\it Phys. Lett.} {\bf B 134}, 194, (1984).

\bibitem{CW}
M. Cahen and N. Wallach,
{\it Bull. Am. Math. Soc.} {\bf 76},  585 (1970).

\bibitem{KSMH}
D. Kramer, H. Stephani, M. MacCallum, and E. Herlt, {\it Exact
Solutions of the Einstein Field Equations\/}, CUP,
Cambridge, 1980.

\bibitem{FP}
 J. Figueroa-O'Farrill and G. Papadopoulos,
 Homogeneous fluxes, branes and a maximally supersymmetric solution of
 M-theory,  {\it JHEP} {\bf 0108}, 036 (2001);
 hep-th/0105308.

\bibitem{BFHP}
 M. Blau, J. Figueroa-O'Farrill, C. Hull and
 G. Papadopoulos,
 A new maximally supersymmetric background of IIB
 superstring theory,
 {\it JHEP} {\bf 0201}, 047 (2002); hep-th/0110242.

\bibitem{BFHP:0201}
 M. Blau, J. Figueroa-O'Farrill, C. Hull and
 G. Papadopoulos,
  Penrose limits and maximal supersymmetry,
  {\it Class. Quantum Grav.}  {\bf 19}, L87 (2002);
  hep-th/0201081.

\bibitem{BFP}
M. Blau, J. Figueroa-O'Farrill and G. Papadopoulos,
Penrose limits, supergravity and brane dynamics,
{\it Class. Quantum Grav.} {\bf 19}, 4753 (2002);
hep-th/0202111.

\bibitem{Pen}
R. Penrose.
In M. Cahen and M. Flato, editors,
Differential geometry and
relativity, page 271. D. Reidel, Dordrecht, 1976.

\bibitem{Guv}
R. G\"{u}ven.
{\it Phys. Lett.} {\bf B 482}, 255 (2000);
hep-th/0005061.


\bibitem{Me}
R.R. Metsaev,
Type IIB Green-Schwarz superstring in plane wave Ramond-Ramond
background,
{\it Nucl. Phys.} {\bf B 625}, 70 (2002); hep-th/0112044.

\bibitem{MT}
R.R. Metsaev and A.A. Tseytlin,
Exactly solvable model of superstring in plane wave Ramond-Ramond
background,
{\it Phys. Rev.} {\bf D 65}, 126004 (2002); hep-th/0202109.

\bibitem{RT}
J.G. Russo and A.A. Tseytlin,
On solvable models of type IIB superstring in NS-NS
and R-R plane wave  backgrounds,
{\it JHEP} {\bf 0204}, 021 (2002); hep-th/0202179.

\bibitem{BMN}
D. Berenstein, J. Maldacena and H. Nastase,
Strings in flat space and pp-waves from N = 4 super Yang Mills,
{\it JHEP} {\bf 0204}, 013 (2002); hep-th/0202021.

\bibitem{MM}
J. Maldacena and L. Maoz, String on $pp$-waves and massive
two dimensional field theories, hep-th/0207284.

\bibitem{BGS}
D. Brecher, J.P. Gregory and P.M. Saffin,
String theory and the classical stability of plane waves,
hep-th/0210308.

\bibitem{MR}
D. Marolf and S. F. Ross, On the singularity structure and
stability of plane waves, hep-th/0210309.

\bibitem{FP1}
J. Figueroa-O'Farrill and  G. Papadopoulos,
Maximally supersymmetric solutions of ten- and eleven-dimensional
supergravities,  hep-th/0211089.

\bibitem{PRT}
G. Papadopoulos, J.G. Russo and A.A. Tseytlin, Solvable model
of strings in a time-dependent plane-wave background,
hep-th/0211289.

\bibitem{NKim}
N. Kim, Comments on $IIB$ $pp$-waves with Ramaond-Ramond
fluxes and massive two dimensional non-linear sigma-models,
hep-th/0212017.

\bibitem{CMPPPZ}
A. Coley, R. Milson, N. Pelavas, V. Pravda, A. Pravdova
and R. Zalaletdinov, Generalized $pp$-wave spacetimes in higher
dimensions, gr-qc/0212063.

\bibitem{BL}
M. Blau and M. O'Loughlin,
Homogeneous plane waves, hep-th/0112135.

\bibitem{Ahm}
E.T. Ahmedov, On the relations between correlation functions in
$SYM/pp$-wave correspondence, hep-th/0212297.

\bibitem{CJS}
E. Cremmer, B. Julia, and J. Scherk,
{\it Phys. Lett.} {\bf B76}, 409 (1978).

\bibitem{Reid}
J.L. Reid, An exact solution of the non-linear
differential equation $\ddot{y} + p(t) y =
q_m(t)/y^{2m -1}$. {\it Proc. Amer. Math. Soc.}
{\bf 27}, 61 (1971).

\bibitem{IMtop}
V.D. Ivashchuk and V.N. Melnikov,  Exact solutions in
multidimensional gravity with antisymmetric forms,
topical review, {\it Class. Quantum Grav.} {\bf 18},
R82-R157 (2001); hep-th/0110274.

\bibitem{AR}
I.Ya. Aref'eva and O.A. Rytchkov,
Incidence matrix description of intersecting p-brane
solutions, {\it Preprint} SMI-25-96, hep-th/9612236.

\bibitem{IM3}
V.D. Ivashchuk and V.N. Melnikov,
Sigma-model for the generalized  composite p-branes,
{\it  Class. Quantum Grav.} {\bf 14}, 3001 (1997); hep-th/9705036.

\bibitem{I2}
V.D. Ivashchuk, Composite p-branes on product of Einstein spaces,
{\it Phys. Lett.} {\bf B 434}, 28-35, (1998); hep-th/9704113.

\bibitem{KKLP}
N. Khvengia, Z. Khvengia, H. L\"u and  C.N. Pope,
Toward field theory of F-theory, hep-th/9703012.

\bibitem{F-th}
C.M. Hull, String dynamics at strong coupling
{\it Nucl. Phys.} {\bf B 468}, 113 (1996); \\
C. Vafa, Evidence for F-Theory,
{\it Nucl. Phys.} {\bf B 469}, 403 (1996);
hep-th/9602022.

\bibitem{IMJ}
V.D. Ivashchuk and V.N. Melnikov,
Multidimensional classical and quantum cosmology
with intersecting $p$-branes,
{\it J. Math. Phys.} {\bf 39}, 2866-2889 (1998);
hep-th/9708157.

\bibitem{IMb}
V.D. Ivashchuk and V.N. Melnikov, Billiard representation for multi-
dimensional cosmology with intersecting p-branes near the singularity.
{\it J. Math. Phys.} {\bf 41}, 6341-6363 (2000);
hep-th/9904077.

\bibitem{DHN}
T. Damour, M. Henneaux and H. Nicolai,
Cosmological billiards, topical review, to appear in {\it Class. Quantum
Grav.} (2003); hep-th/0212256.

\bibitem{Is-b}
V.D. Ivashchuk, Composite S-brane solutions
related to Toda-type systems, {\it Class. Quantum
Grav.} {\bf 20}, 261-276 (2003); hep-th/0208101.

\bibitem{Isusy}
V.D. Ivashchuk, On supersymmetric solutions
in $D = 11$ supergravity on product of Ricci-flat
spaces, {\it Grav.  Cosmol. } {\bf 6}, No 4 (24), 344-350 (2000);
hep-th/0012263.


\end{thebibliography}
\end{document}